\documentclass[a4paper,lnbip]{svmultln}
\usepackage{graphicx}
\usepackage{latexsym}
\usepackage{float}
\usepackage{amssymb}
\usepackage{amsmath}
\usepackage{tabularx}
\usepackage{wrapfig}

\graphicspath{{img/}}

\begin{document}
\mainmatter 
 
\title{Making Sense of Declarative Process Models: Common Strategies and Typical
Pitfalls\thanks{This research is supported by Austrian Science Fund (FWF):
P23699-N23 and the BIT fellowship program. The final publication is available at
Springer via http://dx.doi.org/10.1007/978-3-642-38484-4\_2}}
\titlerunning{Making Sense of Declarative Process Models}
\author{Cornelia Haisjackl\inst{1} \and  Stefan Zugal\inst{1} \and Pnina
Soffer\inst{2} \and Irit Hadar\inst{2} \and \\Manfred Reichert\inst{3} \and
Jakob Pinggera\inst{1} \and Barbara Weber\inst{1}}
\authorrunning{Haisjackl et al.}

\institute{University of Innsbruck, Austria\\
\email{{cornelia.haisjackl, stefan.zugal, jakob.pinggera,
barbara.weber}@uibk.ac.at} \and University of Haifa, Israel\\ \email{{spnina,
hadari}@is.haifa.ac.il} \and University of Ulm, Germany\\
\email{manfred.reichert@uni-ulm.de}
}
\maketitle

\vspace{-0.4cm}
\begin{abstract}
Declarative approaches to process modeling are regarded as well suited for
highly volatile environments as they provide a high degree of flexibility. However,
problems in understanding and maintaining declarative business process models
impede often their usage. In particular, how declarative models are understood
has not been investigated yet. This paper takes a first step toward
addressing this question and reports on an exploratory study
investigating how analysts make sense of declarative process models.
We have handed out real-world declarative process models to
subjects and asked them to describe the illustrated process. Our qualitative
analysis shows that subjects tried to describe the processes in a
\textit{sequential way} although the models represent circumstantial
information, namely, conditions that produce an outcome, rather than a sequence
of activities. Finally, we observed difficulties with single building
blocks and combinations of relations between activities.

\end{abstract}

\vspace{-1.0cm}
\section{Introduction}
\label{sec:introduction}
Regarding the analysis and design of information systems, conceptual modeling has
proven to foster understanding and communication~\cite{Mylo98}. For example,
\textit{business process models} (\textit{process models} for short) have been
employed in the context of process-aware information systems, service-oriented
architectures, and web services~\cite{ReMe}. Recently, \textit{declarative}
process models have gained attention due to their flexibility with respect to
modeling and execution of processes~\cite{ReWe12}. While technical issues of
declarative process modeling, such as formalization of semantics~\cite{HiMS12},
maintainability \cite{ZuPW11}, verification \cite{Pesi08}, and execution
\cite{Barba_2013} are well understood, understandability issues of declarative
models have not been investigated in detail yet. In particular, it has been
argued that understandability may be hampered by lack of computational
offloading~\cite{Zug+11a} or hidden dependencies~\cite{ZuPW12}. Put differently,
it is not entirely clear whether the full potential of declarative modeling can
be exploited or whether understandability issues will interfere.

We approach these issues by studying the sense-making of declarative process
models through the lens of an empirical investigation. In particular, we handed
out declarative process models to subjects and asked them to describe the
illustrated process. In addition, we asked them to voice their thoughts while
describing the process, i.e., we applied \textit{think-aloud techniques}
\cite{ErSi93} to get insights into the subject's reasoning processes. Since we
were interested in how different structures of the process representation would
influence process model understandability, we maintained another variant of each
model describing the same process, but making use of modularization, i.e.,
sub-processes. The contribution of this work is twofold. On one hand, we provide
insights into how subjects make sense of declarative process models, e.g., we
analyze strategies how to read declarative process models. On the other, we
consider characteristic problems that occur when scanning declarative process
models. Our contribution aims at guiding the future development of supporting
tools for system analysts, as well as pointing out typical pitfalls to teachers
and educators of analysts.

The exploratory study reported in this paper is part of a larger investigation on
declarative process models. While our previous work focused on quantitative
results, this paper deals with qualitative data solely.\footnote{The
exploratory study's material can be downloaded from:\\
http://bpm.q-e.at/experiment/HierarchyDeclarative} Sect.~\ref{sec:background}
gives background information. Sect.~\ref{sec:methodology} describes the
setup of the exploratory study, whereas Sect.~\ref{sec:case_study} deals with its
execution. Sect.~\ref{sec:results} presents the results of the exploratory study
and Sect.~\ref{sec:discussion} a corresponding discussion. Related work is
presented in Sect.~\ref{sec:related_work}. Finally, Sect.~\ref{sec:summary}
concludes the paper.

\vspace{-0.4cm}
\section{Background: Declarative Process Models}
\label{sec:background} 
There has been a long tradition of modeling business processes in an imperative
way. Process modeling languages supporting this paradigm, like BPMN and EPC, are
widely used. Recently, \textit{declarative approaches} have received increasing
interest, as they suggest a fundamentally different way of describing business
processes~\cite{Pesi08}. While imperative models specify exactly \textit{how}
things must be done, declarative approaches focus on the logic that governs the
interplay of process actions by describing
\textit{activities} that may be performed, as well as \textit{constraints}
prohibiting undesired behavior. Constraints found in literature may be divided
into existence constraints, relation constraints, and negation constraints
\cite{AalstPesic2006DecSerFlow}. \textit{Existence constraints} specify how often
an activity must be executed for one particular process instance. In
turn, \textit{relation constraints} restrict the ordering of activities by
imposing respective restrictions. Finally, \textit{negation constraints} define
negative relations between activities. Table~\ref{tab:constraintsDef} shows
examples for each category, an overview of all constraints can be found
in~\cite{AalstPesic2006DecSerFlow}.

\begin{table}[htb]
        \centering
        \begin{tabularx}{120mm}{l|l|X}
        \hline
        \textbf{Group}&\textbf{Constraint}&\textbf{Definition}\\\hline
        existence & exactly(a,n)&activity $a$ must occur exactly $n$ times\\\hline
        & existence(a,n)&$a$ must occur at least $n$ times\\\hline
        & init(a)&$a$ must be the first executed activity in every trace\\\hline
        & last(a)&$a$ must be the last executed activity in every trace\\\hline\hline
        relation  & precedence(a,b)&activity $b$ must be preceded by activity
        $a$ (but not necessarily directly preceded)\\\hline
        & response(a,b)&if $a$ is executed, $b$ must be executed afterwards (but
        not necessarily directly afterwards)\\\hline
        & chain\_response(a,b)&if $a$ is executed, $b$ is executed directly
        afterwards\\\hline
        & coexistence(a,b)&if $a$ is executed, $b$ must be executed and
        vice-versa\\\hline\hline
        negation  & neg\_response(a,b)&if $a$ is executed, $b$ must not be
        executed afterwards\\\hline & neg\_coexistence(a,b)&$a$ and $b$
        cannot co-occur in any trace\\\hline
        \end{tabularx}
    \caption{Definition of constraints}
    \label{tab:constraintsDef}
    \vspace{-0.5cm}
\end{table}

\begin{figure}[htp]
 \centering
 \includegraphics[width=.85\textwidth]{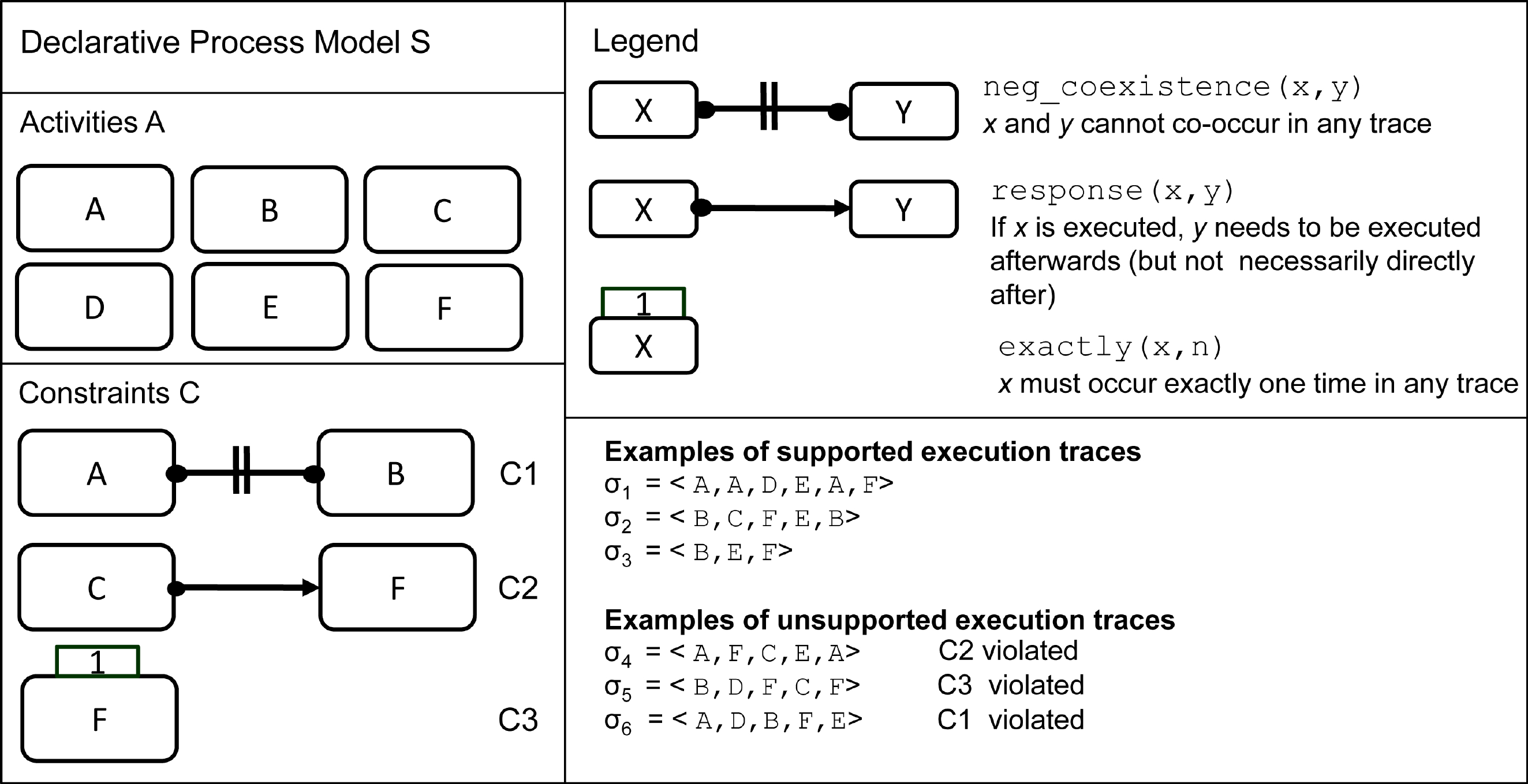}
 \caption{Example of a declarative process model}
 \label{fig:example_declModel}
 \vspace{-0.7cm}
\end{figure}

An example of a declarative process model $S$ specified with ConDec~\cite{Pesi08}
is depicted in \figurename~\ref{fig:example_declModel}. The model consists of six
distinct activities \texttt{A}, \texttt{B}, \texttt{C}, \texttt{D}, \texttt{E},
and \texttt{F}. In addition, it comprises three constraints. The
\textit{neg\_coexistence} constraint, i.e., C1, forbids that \texttt{A} and
\texttt{B} co-occur in the same trace. In turn, the \textit{response}
constraint, i.e., C2, requires that every execution of \texttt{C} must be
followed by one of \texttt{F} before the process instance may complete. Finally,
the \textit{exactly} constraint, i.e., C3, states that \texttt{F} must be
executed exactly once per process instance. While instances with traces \texttt
{$\sigma_1$=<A,A,D,E,A,F>}, \texttt {$\sigma_2$=<B,C,F,E,B>}, and
\texttt{$\sigma_3$=<B,E,F>} satisfy all the constraints,
\texttt{$\sigma_4$=<A,F,C,E,A>} violates C2, \texttt{$\sigma_5$=<B,D,F,C,F>}
violates C3, and \texttt {$\sigma_6$=<A,D,B,F,E>} violates C1.
\texttt{$\sigma_5$=<B,D,F,C,F>} highlights a \textit{hidden dependency} between
\texttt{C} and \texttt{F}. The combination of the \textit{exactly}
constraint, i.e., C3, and the \textit{response} constraint, i.e., C2, adds an
implicit constraint that does not exist when looking at the constraints in
isolation. This hidden dependency prohibits that \texttt{F} is executed before
\texttt{C}, assuming that \texttt{C} is executed at all.

\vspace{-0.2cm}

\subsubsection{Hierarchy in Declarative Process Models.} 
\begin{figure}[htp]
 \centering
 \includegraphics[width=.60\textwidth]{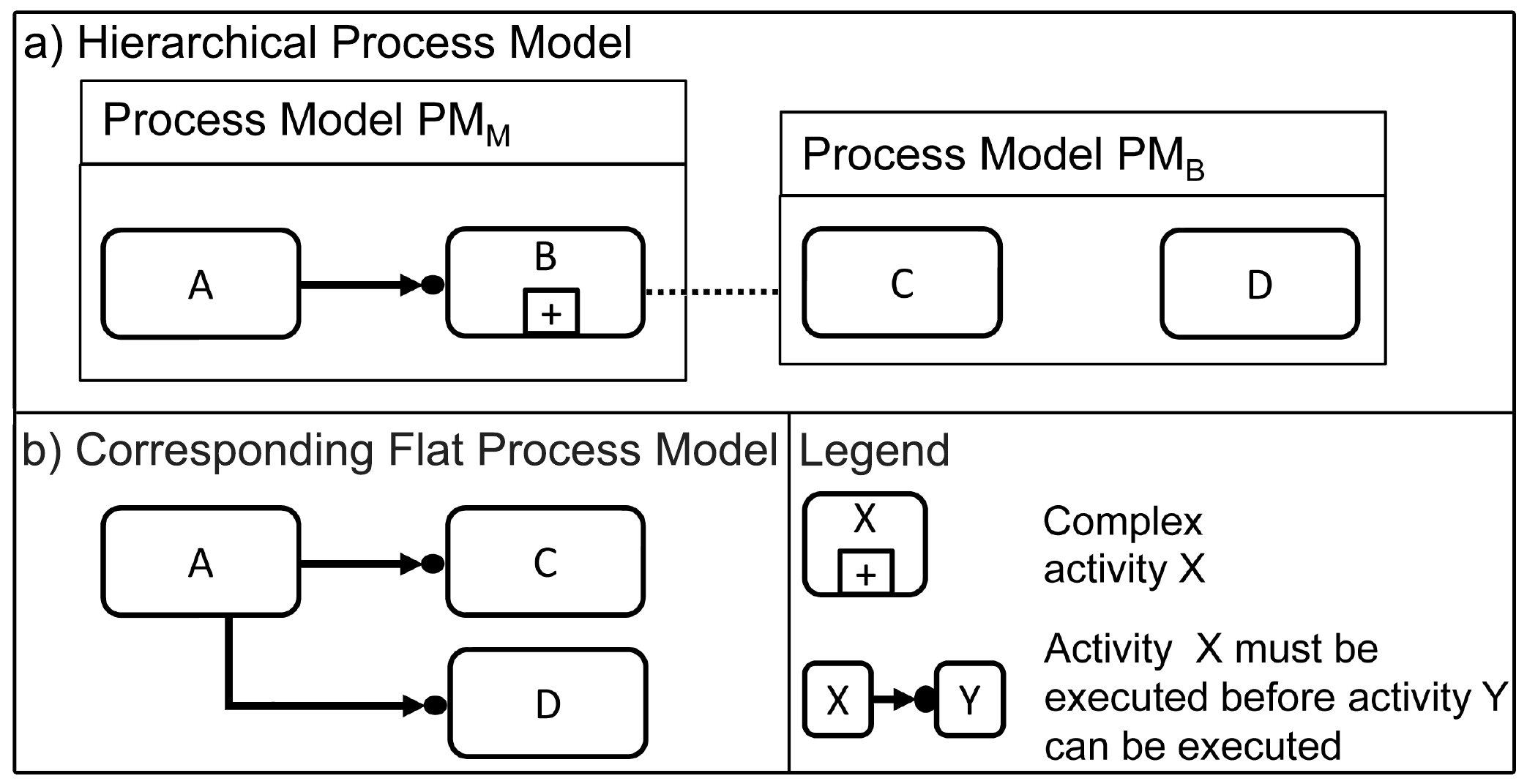}
 \caption{Example of a process model with and without hierarchy}
 \label{fig:example_hier}
\end{figure}

Using modularization to
hierarchically structure information has been identified as a viable approach to
deal with complexity for decades~\cite{Parn72}. Taking a look at declarative
process models with hierarchy in general, a sub-process may be introduced in a
process model via a \textit{complex activity}, referring to a process model. When
the complex activity is executed, the referred process model, i.e., the
sub-process, is instantiated (see \cite{Zug+12a} for details).
\figurename~\ref{fig:example_hier}a) shows a hierarchical model, complex activity
$B$ refers to a sub-process that contains activities $C$ and $D$.
\figurename~\ref{fig:example_hier}b) shows the corresponding flat process model.
Even though \figurename~\ref{fig:example_hier}a) and
\figurename~\ref{fig:example_hier}b) are semantically equivalent, they differ in
the number of activities and constraints.

\vspace{-0.3cm}

\vspace{-1cm}
\section{Defining and Planning the Exploratory Study}
\label{sec:methodology}
In order to investigate how subjects make sense of declarative process models we
conduct an exploratory study. In particular, we are interested in common
strategies and typical pitfalls occurring during this sense-making process. Since
there has been no considerable research on understandability issues of
declarative process models, and hence no theories exist we can base our
investigation on, we address the topic in an exploratory manner using a
qualitative research approach~\cite{bassey1999case}. In particular, we use the
think-aloud method, i.e., we ask participating subjects to voice their thoughts,
allowing for a detailed analysis of their reasoning process~\cite{ErSi93}. Then,
we turn to grounded theory~\cite{corbin2007basics}, an analysis approach for
identifying recurring aspects and grouping them to categories. These categories
are validated and refined throughout the analysis process. First of all, we
describe setup and planning of the exploratory study.
% cite: 41,45,46 from Irit's Paper (when intuition and logic clash)
% 46: bogdan1998qualitative

\subsubsection{Subjects.} In order to ensure that obtained results are
not influenced by unfamiliarity with declarative process modeling, subjects need
to be sufficiently trained. Even though we do not require experts, subjects should
have at least a moderate understanding of declarative processes' principles.

\subsubsection{Objects.} The process models used in the study originate from a
case study~\cite{Hais12} and describe real-world business processes. From a set
of 24 process models collected in this case study, 4 models were chosen as basic
\textit{objects} for the exploratory study. This was accomplished in a way
ensuring that the numbers of activities and constraints vary. To make the models
amenable for this study, they underwent the following procedure. First, the
models were translated to English (the case study was conducted in German) since
all exercises were done in English (four subjects did not speak German). Second,
since the models collected during the modeling sessions had not gone through
quality assessment, they were scanned for errors and corrected accordingly.
Third, since we were interested in how different structures of the process
representation would influence the process models' understandability, we created
a second variant of each process describing the same process, but making use of
sub-processes. Consequently, we have two variants of each process model: a flat
and a hierarchical one.
\vspace{-0.3cm}
\begin{table}[htb]
        \centering
        \begin{tabular}{|l|l|c|c|c|c|}
        \cline{2-6}
        \multicolumn{1}{c|}{}&\multicolumn{1}{c|}{\textbf{Type}}&\textbf{Proc.
        1}&\textbf{Proc. 2}& \textbf{Proc. 3}& \textbf{Proc. 4}\\
        \hline
        Activities       & flat      & 11 & 8 & 23 & 23\\ \cline{2-6}
                         & hierarchy & 13 & 9 & 26 & 26\\ \hline
        Constraints      & flat      & 19 & 7 & 30 & 45\\ \cline{2-6}
                         & hierarchy & 21 & 9 & 28 & 44\\ \hline
        Constr. types    &           & 8  & 4 & 7  & 5 \\ \hline
        Sub-processes    & hierarchy & 2  & 1 & 3  & 2 \\ \hline
        Components       &           & 2  & 5 & 2  & 2\\  \hline
        Domain           & & Software & Teaching & Electronic & Buying an\\ 
                         & & development & & company & apartment\\
                         \hline
        \end{tabular} 
    \caption{Characteristics of the process models used in this study}
    \label{tab:models}
    \vspace{-0.5cm}
\end{table}

Table~\ref{tab:models} summarizes the characteristics of the process models. The
latter comprise between 8 and 26 activities, and between 7 and 45 constraints.
The differences in the number of activities between flat and hierarchical models
are caused by the complex activities representing sub-processes in the
hierarchical models (cf. Sect. \ref{sec:background}). Similarly, constraints had
to be added or removed to preserve the behavior when creating a hierarchical
model. Process models vary regarding the degree of interconnectivity of
constraints, i.e., models consist of two to five components (cf.
Table~\ref{tab:models}). A component is defined as a part of the model where any
two activities are connected by constraints, and not connected to any other
activity in the model. The process models are based on four different domains
describing bug fixing in a software company, a teacher's preparations prior to
teaching, a worker's duties at an electronic company, and buying and renovating
an apartment (cf. Table~\ref{tab:models}). The process models contain constraints
of all three types, i.e., existence, relation, and negation constraints, except
the second process model (no negation constraints). Table~\ref{tab:constraints}
provides additional information on the constraint types included in each process
model.
%http://en.wikipedia.org/wiki/Connected_component_(graph_theory)
% In graph theory, a connected component of an undirected graph is a subgraph in
% which any two vertices are connected to each other by paths, and which is
% connected to no additional vertices in the supergraph.

\begin{table}[htb]
        \centering
        \begin{tabular}{|l|l||c|c|c|c||c|c|c|c|}
        \cline{3-10}
        \multicolumn{2}{c|}{}&\multicolumn{4}{c||}{\textbf{flat}}&
        \multicolumn{4}{c|}{\textbf{hierarchical}}\\ \hline
        \textbf{Group}&\textbf{Constraint}&\textbf{P 1}&\textbf{P 2}&
        \textbf{P 3}& \textbf{P 4}&\textbf{P 1}&\textbf{P 2}&
        \textbf{P 3}& \textbf{P 4}\\
        \hline \hline
        existence & existence constraints&5&2&1&10&7&4&1&13\\\hline
                  & init                 &1&1&1&0&1&1&1&0\\\hline
                  & last                 &0&1&0&1&0&1&0&1\\\hline \hline  
        relation  & precedence           &4&3&18&20&4&3&18&20\\\hline
                  & response             &0&0&1&0&0&0&1&0\\\hline
                  & succession           &0&0&0&4&0&0&0&4\\\hline
                  & coexistence          &0&0&1&0&0&0&1&0\\\hline
                  & chained precedence   &2&0&0&0&2&0&0&0\\\hline
                  & chained response     &3&0&1&0&3&0&1&0\\\hline
                  & chained succession   &1&0&0&0&1&0&0&0\\\hline \hline
        negation  & negation response    &2&0&0&10&2&0&0&6\\\hline
                  & mutual exclusion     &1&0&7&0&1&0&5&0\\\hline
        \end{tabular} 
    \caption{Constraints of the process models used in this study}
    \label{tab:constraints}
    \vspace{-0.5cm}
\end{table}

% \begin{table}[htb]
%         \centering
%         \begin{tabular}{|l|l|c|c|c|c|}
%         \hline
%         \textbf{Group}&\textbf{Constraint}&\textbf{Proc. 1}&\textbf{Proc. 2}&
%         \textbf{Proc. 3}& \textbf{Proc. 4}\\
%         \hline
%         existence & existence constraints& 5 & 2 & 1  & 10\\ \hline
%                   & init                 & 1 & 1 & 1  & 0\\ \hline
%                   & last                 & 0 & 1 & 0  & 1\\ \hline \hline  
%         relation  & precedence           & 4 & 3 & 18 & 20\\\hline 
%                   & response             & 0 & 0 & 1  & 0\\ \hline
%                   & succession           & 0 & 0 & 0  & 4\\ \hline
%                   & coexistence          & 0 & 0 & 1  & 0\\ \hline
%                   & chained precedence   & 2 & 0 & 0  & 0\\ \hline 
%                   & chained response     & 3 & 0 & 1  & 0\\ \hline
%                   & chained succession   & 1 & 0 & 0  & 0\\ \hline \hline
%         negation  & negation response    & 2 & 0 & 0  & 10\\ \hline
%                   & mutual exclusion     & 1 & 0 & 7  & 0\\ \hline
%         \end{tabular} 
%     \caption{Constraints of the flat process models used in this study}
%     \label{tab:constraints}
% \end{table}

\subsubsection{Design.} 
\begin{figure}
\begin{center}
  \includegraphics[width=\textwidth]{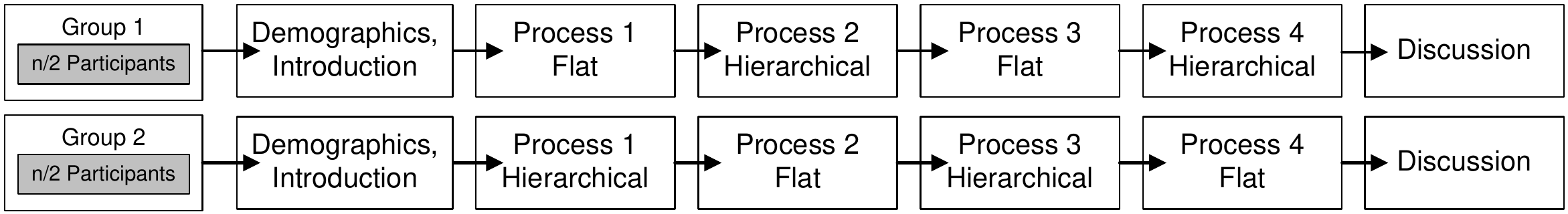}
  \caption{Design of the exploratory study}
  \label{fig:experimental_design}
  \vspace{-1cm}
\end{center}
\end{figure}

\figurename~\ref{fig:experimental_design} shows the
overall design of the exploratory study: First, subjects are \textit{randomly}
assigned to two groups of similar size. Regardless of the group assignment,
demographical data is collected and subjects obtain introductory assignments. To
support subjects in their task, cheat sheets briefly summarizing the constraints'
semantics, are provided, which can be used throughout the study. Introductory
tasks allow subjects to familiarize themselves with the type of tasks to be
performed---potential problems can therefore be resolved at this stage without
influencing actual data collection. After this familiarization phase, subjects
are confronted with the actual tasks. Each subject works on two flat process
models and two hierarchical ones. Group 1 starts with the flat representation of
process model 1, while Group 2 works on the hierarchical representation of the
same model. Subjects are confronted with hierarchical and flat models in an
alternating manner. For each model, the subject is asked to \textit{``explain
roughly what the process describes.''} The exploratory study is concluded by a
discussion with the subject to help reflecting on the study and providing us with
feedback.

\vspace{-0.2cm}

\subsubsection {Instrumentation.} For each model, subjects received separate
paper sheets showing the process models, allowing them to use a pencil for
highlighting or taking notes, and juxtaposing the process models as desired. No
written answers were required, only free talking. Audio and video recording are
used as it has proven being useful for resolving unclear situations in
think-aloud protocols \cite{Zug+12c}.

%Zitat Irit: The multiple sources of the data collected, along with the three
% different workshop settings, enabled triangulation, thus increasing the
% validity of the findings.

\vspace{-0.5cm}
\section{Performing the Exploratory Study}
\label{sec:case_study}
%This section describes how the exploratory study was performed.

\subsubsection{Execution.} The study was conducted in July 2012 in two locations.
First, seven subjects participated at the University of Ulm, followed by two
additional sessions at the University of Innsbruck, i.e., a total of nine
subjects participated. To ensure that subjects were sufficiently familiar with
declarative process modeling, they were provided with training material. Each
session was organized as follows: First, the subject was welcomed and instructed
to speak thoughts out loudly. To allow subjects to concentrate on their tasks,
the sessions were performed in a ``paper-workflow'' manner, i.e., one supervisor
was seated left to the subject, a second supervisor to the right. The sheets
containing the study's material were then passed from the left to the subject. As
soon as the subject finished the task, the material was passed to the supervisor
on the right. Meanwhile, the subject's actions were audio- and video-recorded to
gather any uttered thoughts.

\subsubsection{Data Validation.} In each session, only a single subject
participated, allowing us to ensure that the study setup was obeyed. In addition,
we screened whether subjects fitted the targeted profile, i.e., were familiar
with process modeling and ConDec~\cite{Pesi08}. We asked questions regarding
familiarity on process modeling, ConDec, and domain knowledge; note that the
latter may significantly influence performance~\cite{Kha+06}. For this, we
utilize a 7-point Likert Scale, ranging from \textit{``Strongly agree'' (7)} over
\textit{``Neutral'' (4)} to \textit{``Strongly disagree'' (1)}. Results are
summarized in Table~\ref{tab:demographics}. Finally, we assessed the subjects'
professional background: all subjects indicated an academic background, i.e.,
were either PhD students or postdocs. We conclude that they had a profound
background in process modeling (the least experienced subject had 2.5 years of
modeling experience) and were moderately familiar with ConDec.

\vspace{-0.3cm}
\begin{table}[ht]
     \centering
     \begin{tabular}{|l|c|c|c|}
            \hline & Minimum & Maximum & Median \\ \hline
            1) Years of modeling experience & 2.5 & 7 & 5\\ 
            2) Models read last year & 10 & 250 & 40\\ 
            3) Models created last year & 5 & 100 & 10\\
            4) Average number of activities & 5 & 50 & 15\\ \hline 
            5) Familiarity ConDec & 2 & 6 & 3\\
            6) Confidence understanding ConDec & 2 & 6 & 4\\
            7) Confidence creating ConDec & 2 & 6 & 4\\ \hline
            8) Familiarity software development & 4 & 7 & 6\\
            9) Familiarity teaching & 4 & 7 & 5\\
            10) Familiarity electronic companies & 1 & 6 & 2\\
            11) Familiarity buying apartments & 1 & 6 & 4\\ \hline
    \end{tabular}
    \vspace{5pt} 
    \caption{Demographics (5--11 based on 7-point Likert Scale)}
    \label{tab:demographics}
    \vspace{-0.9cm}
\end{table}

\subsubsection{Data Analysis.} Our research focuses on sense-making of
declarative process models. On one hand, we investigate strategies applied by
subjects in understanding process models, on the other, we explore typical
phenomena and pitfalls in this process. For this purpose, data analysis
comprised the following stages.

\begin{enumerate}
  \item Transcription of the subjects' verbal utterances
  \item Creation of graphs describing the order in which subjects mention
  activities
  \item Analysis of transcripts using grounded theory
\end{enumerate}

In (2), for each process model we create a graph representing the order
activities were mentioned by the subjects. For this purpose, we utilize the
transcripts created in (1), but also video recordings to identify when subjects
visited an activity without talking about it. In (3), we apply grounded theory to
the transcripts to explore and understand  phenomena appearing when subjects make
sense of declarative process models. As a starting point, transcripts are
inspected, marking aspects that caused confusion, were misinterpreted or left
out. In a second iteration, we revisit the marked areas and search for new
aspects. This process of open coding analysis is repeated until no new aspects
can be found. Afterwards, we perform axial coding, i.e., we repeatedly group
aspects to form high level categories. We count the number of identified markings
per category.

% Axial coding in Grounded Theory is the process of relating codes (categories
% and concepts) to each other, via a combination of inductive and deductive thinking.

\vspace{-0.5cm}
\section{Findings}
\label{sec:results}
Based on the findings of our data analysis, we identified different ways how
declarative models are read and interpreted.

\vspace{-0.2cm}
\subsection{Reading Declarative Business Process Models}
\label{sec:1}
\begin{figure}[htp]
 \centering
 \includegraphics[width=.90\textwidth]{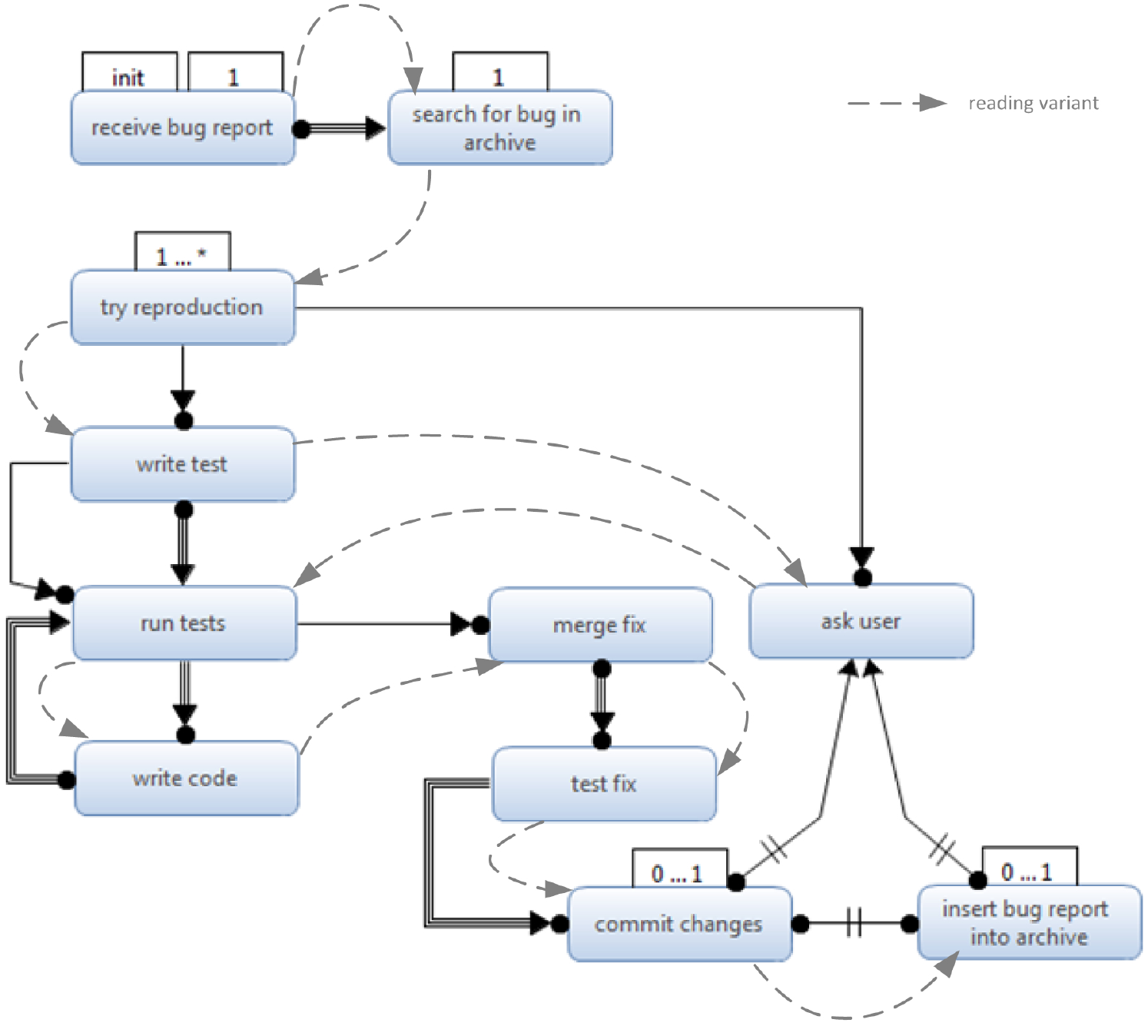}
 \caption{First process model and a reading variant}
 \label{fig:model1ReadingVariant}
 \vspace{-0.3cm}
\end{figure}

When analyzing graphs and transcripts, we observed that subjects consistently
adopted similar strategies when reading declarative models. For example,
\figurename~\ref{fig:model1ReadingVariant} shows the flat version of the first
model and a typical strategy to understand that model. The model consists of two
components. The first one contains activities \textit{``receive bug report''} and
\textit{``search for bug in archive''}. The second component comprises all other
activities. The dotted arrows display how three out of five subjects (Group 1)
read the model to understand it.

Regardless of whether sub-processes were present or not, they described the
process in the order activities were supposedly executed, i.e., tried to describe
the process in a \textit{sequential way}. Hence, as a first step, subjects
skimmed over the process model to find an \textit{entry point} where they could
start with describing the (main) process: \textit{``\ldots Ok, this is the first
activity since it has this init constraint\ldots''} Interestingly, subjects
appreciated when a clear starting point for their explanations could be found:
\textit{``\ldots it is nice that we have an init activity, so I can start with
this\ldots''} Relating to the model depicted in
\figurename~\ref{fig:model1ReadingVariant}, subjects started with
\textit{``receive bug report''} because of the init constraint. Then, they
mentioned \textit{``search for bug in archive''}. A declarative process model,
however, does not necessarily have a unique entry point, apparently causing
confusion: \textit{``Well\ldots gosh\ldots I've got no clue where to start in
this model\ldots''} The subjects used two different solutions for this kind of
situation. Either they looked for a last constraint (\textit{``So, we don't have
init, but we have last\ldots''}) or they assumed the upper left corner of the
model to be its entry point (\textit{``Ok\ldots so first of all I have three
initial I would say start activities\ldots''}). After having identified an entry
point, subjects tried to figure out in which order activities are to be executed:
\textit{``After given duties to the apprentices there should come these two
tasks\ldots''}

This routine was iterative, i.e., if parts of a model were not connected,
subjects applied the same strategy for each component, i.e., they started again
at the upper left corner of these components. We observed this behavior
independent of the respective process model or subject. Regarding our example
(cf. \figurename~\ref{fig:model1ReadingVariant}), after describing the first
component, subjects took a look at the second one. As there was no init
constraint, they started in the upper left corner (\textit{``try reproduction''})
and followed the other activities in a sequential way. Two subjects had problems,
since there is no connection between the two parts of the model: \textit{``Ah,
then there is a different process because these are not connected\ldots''}
Likewise, a subject got irritated with single activities that had no connection
to the rest of the model: \textit{``...ah this one, there's no constraint here.
You are just trying to confuse me.''} Finally, subjects indicated where the
process supposedly ends: \textit{``\ldots the process ends with the activity give
lessons\ldots''} When there was no last constraint (cf.
\figurename~\ref{fig:model1ReadingVariant}), subjects stopped describing the
process model after having mentioned all activities of all components.

If a model contained sub-processes (cf. Table~\ref{tab:models}), subjects
preferred talking first about the main process in the above specified way before
describing the sub-processes. When reading sub-processes the subjects used the
same routine as for the main process, except two subjects. One of them described
all and the second subject one out of four sub-processes completely
\textit{backwards}, i.e., following the semantics of precedence constraints,
instead of describing them sequentially.

 \vspace{-0.2cm}
\subsection{Single Building Blocks}
\label{sec:2}

\subsubsection{Flat Declarative Process Models.}
\label{sec:2.1}
In general, when subjects try to make sense of a model, they name activities
and their connections. Sometimes, it happened that subjects missed
single or small groups of activities. Regarding the first and second flat
process model, no activities were left out. Three out of five subjects missed
either 2, 4 and 17 activities in the flat version of the third process
model (26 activities). The 17 activities were not referred to
because a subject did not look at the components of the process model in detail.
Four activities were not mentioned in the fourth flat process model: two out of four
subjects did not mention one and three activities out of 26 activities.
In summary, 27 out of 294 activities were missed in flat process models.

When describing a model sequentially, subjects name activities explicitly and
most of the connections, i.e., the constraints, implicitly. However, most
subjects did not mention existence constraints. This behavior could not be found
for any other constraint. For 12 out of 18 models (9 subjects described two flat
process models) subjects ignored one or more existence constraints.
Table~\ref{tab:existence} shows the number of possible mentions of existence
constraints per process model and the number of existence constraints that were
ignored by the subjects. Summing up, subjects left out 34 of 78 existence
constraints in flat process models.

\vspace{-0.3cm}
\begin{table}[htb]
        \centering
        \begin{tabular}{|l|l|c|c|c|c|}
        \hline
        \textbf{Number of} & \textbf{Type}&\textbf{Proc. 1}&\textbf{Proc. 2}&
        \textbf{Proc. 3}& \textbf{Proc. 4}\\ \hline \hline
        possible mentions of & flat & 25 & 8 & 5 & 40\\ \cline{2-6}
        existence constraints & hierarchy & 28 & 20 & 4 & 65\\ \hline \hline
        not mentioned & flat & 9 & 1 & 3 & 21\\ \cline{2-6}
        existence constraints & hierarchy & 19 & 2 & 1 & 30 \\ \hline
        \end{tabular} 
    \caption{Existence constraints}
    \label{tab:existence}
    \vspace{-0.7cm}
\end{table}

\subsubsection{Hierarchical Declarative Process Models.}
\label{sec:2.2}
Regarding hierarchical process models, subjects tended to miss less activities.
Two out of four subjects forgot to mention one activity in the first process
model (11 activities). Regarding the second and third process model, no
activities were left out. Three out of five subjects missed one activity in the
hierarchical version of the fourth process model (23 activities). In summary, 5
out of 331 activities were missed in hierarchical process models.

Concerning the existence constraints in hierarchical process models, for 11 out
of 18 models (9 subjects described two hierarchical process models), one or more
existence constraints were not mentioned. As shown in Table~\ref{tab:existence},
52 from 117 existence constraints were ignored in hierarchical process models.

\subsubsection{Flat and Hierarchical Declarative Process Models.}
\label{sec:2.3}
As far as the interpretation of constraints is concerned, subjects had relatively
little problems irrespective of whether the models were flat or hierarchical. As
illustrated in Table~\ref{tab:constraints}, 12 different constraint types were
used in the experimental material. To accomplish their task, subjects had cheat
sheets available and could look constraints they did not know up. Except for the
precedence constraint, which caused considerable difficulties, subjects faced no
notable problems. Four out of nine subjects used the precedence constraint in a
wrong way. According to Sect. \ref{sec:background}, the definition of this
constraint is that \textit{``B can only be executed, if A has been executed
before''}. The subjects used it the other way round, i.e., \textit{``So if we
perform receive incoming good [A] then do quality check [B] should be performed
afterwards\ldots''} One subject stated the absence of the precedence constraint
between two components of a model: \textit{``But still I don't get the relation
between this part and the other one, so this is the problem, because I understand
the flow, but I don't understand the relation between the two parts. Because
there is no precedence.''} Additionally, the missing direction of the coexistence
constraint caused one subject troubles: \textit{``Let's see, so I would say you
kinda start with these two activities, I'm not sure which one\ldots''}

\vspace{-0.2cm}
\subsection{Combination of Constraints}
\label{sec:3}
% This section elaborates on the combination of constraints and the
% problems the subjects were facing. There are two kinds of situations.

\subsubsection{Constraints between two Activities.}
\label{sec:3.1}
The first process model contained two and the fourth process model five
situations where two constraints link two activities. In 6 out of these 7 cases,
the direction of the constraint arrows are directly opposed to each other. For
example, one needs to get offers for interior of an apartment before buying them
(precedence constraint). After the interior is bought, it is not reasonable to
get new offers (negation response). The subjects had no troubles to understand
these situations. However, in the first process model there is a case where a
precedence constraint and a chained response constraint link the two activities
\textit{``write test''} and \textit{``run tests''}. Both arrows are pointing to
the second activity (cf. \figurename~\ref{fig:model1ReadingVariant}). The
precedence constraint ensures that before the first execution of \textit{``run
tests''}, \textit{``write test''} must be executed at least once, i.e., it is not
possible to run a test before it was written. The chained response constraint
tells us that \textit{``If A has been executed, B must be executed immediately
afterwards.''}, meaning that after the test was written, it must be run directly
afterwards; 4 out of 9 subjects had troubles with ``the second arrow'', i.e., the
precedence constraint. Two of them claimed that it is redundant (\textit{``This
part is redundant, right?''}), two even thought it is wrong (\textit{``Over this
relation, this is a precedence, so I think this is, ah, this can be removed.''}).
The other 5 subjects ignored the precedence constraint.

% \begin{figure}[htp]
%  \centering
%  \includegraphics{model1tdd01.pdf}
%  \caption{Part of the first process model}
%  \label{fig:model1tdd01}
% \end{figure}

\subsubsection{Hidden Dependencies.}
\label{sec:3.2}
Three out of the four process models contain hidden dependencies (cf. Sect.
\ref{sec:background}). Since these interactions are not \textit{explicitly}
visible, it is not sufficient that the analyst only relies on the information
displayed explicitly, but must carefully examine the process model for these
hidden dependencies as well. Our results show that the subjects mostly ignored
hidden dependencies, i.e., only in 8 out of 36 models, a hidden dependency was
mentioned or found: \textit{``I have to execute prepare lesson in detail at least
once, therefore, to fulfill the precedence constraint, I must execute prepare
teaching sequence too.''}
% \textit{``Buy apartment needs to be executed once, so look for and choose and
% apartment also needs to be executed at least once and before buy an
% apartment\ldots''}

\vspace{-0.5cm}
\section{Discussion}
\label{sec:discussion}

\subsubsection{Reading Declarative Process Models.}
Subjects preferred describing process models in an iterative and sequential way.
They started with the entry point of a component describing it in a sequential
way and repeating this procedure for every component of the process model. The
sequential way of describing models is surprising, as it is known that
declarative process models rather convey circumstantial information (overall
conditions that produce an outcome) than sequential information (how the outcome
is achieved) \cite{Fah+09a}. In other words, in an imperative model,
sequences are made explicit, e.g., through sequence flows. In a declarative
process model, however, such information might not be available at all. As
subjects tend to talk about declarative models in a sequential manner, it appears
as if they prefer this kind of information. Interestingly, similar observations
could be made in a case study on declarative process modeling~\cite{Zug+12c}.
Therein, sequential information, such as \textit{``A before B''} or
\textit{``then C''} was preferred for communication.

\subsubsection{Single Building Blocks.}
Regarding the interpretation of single building blocks, subjects mentioned
activities and constraints when trying to understand the model. Overall, they had
relatively little problems with the interpretation of single building blocks.
Exceptions seem to be precedence and existence constraints. As a possible
explanation these constraints are too simple and are thus not mentioned at all;
further, cheat sheets are not used (cf. dual-process theory \cite{Kahneman_2002}
describing the interplay of implicit unconscious and explicit controlled
processes). Another explanation is that subjects were biased by previous
knowledge about imperative models. Regarding the precedence constraint, it nearly
looks like the arrow used in imperative process modeling notations.

\subsubsection{Combining Constraints.}
The interplay of constraints seems to pose a challenge, especially hidden
dependencies. One explanation could be that subjects simply forgot looking for
them, as reading declarative models can quickly become too complex for humans to
deal with~\cite{Pesi08}. As mentioned earlier, in 8 out of 36 models subjects
found a hidden dependency. In 5 of these 8 cases, they were found in the second
process model, which has the smallest number of activities, constraints and
constraint types (cf. Table~\ref{tab:models}). This indicates that if a model is
not too complex, subjects will be able to find hidden dependencies. Given this
finding, it seems plausible that the \textit{automated} interpretation of
constraints can lead to significant improvements regarding the understandability
of declarative process models~\cite{Zug+12b,ZuPW11}.
\subsubsection{Differences between Flat and Hierarchical Process Models.}
Subjects did not distinguish between flat and hierarchical process models when
reading the models. They used the same description strategy for components and
sub-processes. Interestingly, subjects left out more activities in flat than in
hierarchical process models (cf. Sect. \ref{sec:2}). A reason for this phenomenon
could be \textit{abstraction} \cite{Zug+11}, i.e., hierarchy allows aggregating
model information by hiding the internals of a sub-process using a complex
activity. Thereby, information can be easier perceived. All other aspects we
found could be observed in flat and hierarchical models equally.

\subsubsection{Limitations.}
Our work has the following limitations. First, the number of subjects in the
exploratory study is relatively low (9 subjects), hampering result
generalization. Nevertheless, it is noteworthy that the sample size is not
unusual for this kind of empirical investigation due to the substantial effort to
be invested per subject \cite{costain2007cognitive}. Second, even though process
models used in this study vary in the number of activities, number of
constraints, and existence of sub-processes, it remains unclear whether results
are applicable to declarative process models in general, e.g., more complex 
models. Third, all participating subjects indicated academic background,
limiting result generalization. However, subjects indicated profound background
in business process management, hence, we argue that they can be interpreted as
proxies for professionals.

\vspace{-0.5cm}
\section{Related Work}
\label{sec:related_work}
We investigated the sense-making of declarative process models. The understanding
of a declarative process model with respect to modularization has been
investigated in~\cite{Zug+12a}. However, opposed to our work, theory rather than
empirical data is used for analysis. The role of understanding declarative
process models during \textit{modeling} has been investigated in~\cite{ZuPW12}.
Similar to our work, it has been postulated that declarative models are most
beneficial when sequential information is directly available, as empirically
validated in~\cite{Zug+12c,ZuPW11}. With respect to the understanding of process
models in general, work dealing with the understandability of \textit{imperative}
business process models is related. The Guidelines of Modeling (GoM) describe
various quality considerations for process models~\cite{Bec+00}. The so-called
`Seven Process Modeling Guidelines' (7PMG) accumulate the insights from various
empirical studies, e.g.,~\cite{MVDA+08}, to develop a set of actions a
system analyst may want to undertake to avoid issues with respect to
understandability~\cite{MeRA10}. The understandability of imperative process
models is investigated empirically in~\cite{ReMe}. As example of
understandability issues in conceptual systems, \cite{BJM06} investigates if UML
analysis diagrams increase system analysts' understanding of a domain.

The impact of hierarchy on understandability has been studied in various
conceptual modeling languages, such as imperative business process
models~\cite{Rej+11}, ER diagrams~\cite{Mood04}, and UML statechart
diagrams~\cite{Cru+07} (an overview is presented in~\cite{Zug+11}). Still, none
of these works deals with the impact of hierarchy on understandability in
declarative process models.

While the effectiveness and usability of design guidelines for multiple diagrams
were evaluated in \cite{KHH00}, there are neither guidelines for designing nor
for easily understanding declarative process models.

\vspace{-0.5cm}
\section{Summary and Outlook}
\label{sec:summary}
Declarative approaches to business process modeling have recently attracted
interest as they provide a high degree of flexibility~\cite{Pesi08}. However,
the increase in flexibility comes at the cost of understandability, and hence might
result in maintainability problems of respective process
models~\cite{Pesi08,Web+09b,ZuPiWe2012Toward}. The presented exploratory study
investigates how subjects make sense of declarative business process models and
provides insights into occurring problems. The results indicate that subjects
read declarative process models in a sequential way. While single constraints
caused only minor problems with exception of the precedence constraint, the
combination of several constraints seems to be more challenging. More
specifically, the subjects of this exploratory study mostly failed to identify
hidden dependencies caused by combinations of constraints.

Even though the data we collected provided first insights into the process of
understanding declarative models, further investigations are needed. Replications
utilizing more complex models seem to be appropriate means for additional
empirical tests. Although the think-aloud protocols already provide a detailed
view on the reasoning processes of an analyst, we plan to employ eye movement
analysis for more detailed analysis. The latter allows identifying areas, the
analyst is focusing on in combination with insights on the required cognitive
effort (similar to process modeling~\cite{Pin+12a}). Based on these insights,
we intend to evolve our work toward empirically founded guidelines enabling better
understandability of declarative process models.

\vspace{-0.5cm}

\bibliographystyle{splncs}
\bibliography{literature}

\begin{thebibliography}{10}

\bibitem{Mylo98}
Mylopoulos, J.:
\newblock Information modeling in the time of the revolution.
\newblock Information Systems \textbf{23} (1998)  127--155

\bibitem{ReMe}
Reijers, H.A., Mendling, J.:
\newblock {A Study into the Factors that Influence the Understandability of
  Business Process Models}.
\newblock SMCA \textbf{41} (2011)  449--462

\bibitem{ReWe12}
Reichert, M., Weber, B.:
\newblock Enabling Flexibility in Process-Aware Information Systems:
  Challenges, Methods, Technologies.
\newblock Springer (2012)

\bibitem{HiMS12}
Hildebrandt, T., R.~Mukkamala, T.S.:
\newblock Nested dynamic condition response graphs.
\newblock In: Proc. FSEN '12. (2012)  343--350

\bibitem{ZuPW11}
Zugal, S., Pinggera, J., Weber, B.:
\newblock The impact of testcases on the maintainability of declarative process
  models.
\newblock In: Proc. BPMDS '11. (2011)  163--177

\bibitem{Pesi08}
Pesic, M.:
\newblock {Constraint-Based Workflow Management Systems: Shifting Control to
  Users}.
\newblock PhD thesis, TU Eindhoven (2008)

\bibitem{Barba_2013}
Barba, I., Weber, B., Valle, C.D., Ramírez, A.J.:
\newblock {User Recommendations for the Optimized Execution of Business
  Processes}.
\newblock DKE (2013)

\bibitem{Zug+11a}
Zugal, S., Pinggera, J., Weber, B.:
\newblock Assessing process models with cognitive psychology.
\newblock In: Proc. EMISA '11. (2011)  177--182

\bibitem{ZuPW12}
Zugal, S., Pinggera, J., Weber, B.:
\newblock {Toward Enhanced Life-Cycle Support for Declarative Processes}.
\newblock Journal of Software: Evolution and Process \textbf{24} (2012)
  285--302

\bibitem{ErSi93}
Ericsson, K.A., Simon, H.A.:
\newblock {Protocol analysis: Verbal reports as data}.
\newblock MIT Press (1993)

\bibitem{AalstPesic2006DecSerFlow}
Aalst, W., Pesic, M.:
\newblock Decserflow: Towards a truly declarative service flow languages.
\newblock Lecture Notes in Computer Science \textbf{4184} (2006)  1--23

\bibitem{Parn72}
Parnas, D.L.:
\newblock {On the Criteria to be Used in Decomposing Systems into Modules}.
\newblock Communications of the ACM \textbf{15} (1972)  1053--1058

\bibitem{Zug+12a}
Zugal, S., Soffer, P., Pinggera, J., Weber, B.:
\newblock {Expressiveness and Understandability Considerations of Hierarchy in
  Declarative Business Process Models}.
\newblock In: Proc. BPMDS '12. (2012)  167--181

\bibitem{bassey1999case}
Bassey, M.:
\newblock Case study research in educational settings.
\newblock Doing qualitative research in educational settings. Open University
  Press (1999)

\bibitem{corbin2007basics}
Corbin, J., Strauss, A.:
\newblock Basics of Qualitative Research: Techniques and Procedures for
  Developing Grounded Theory.
\newblock SAGE Publications (2007)

\bibitem{Hais12}
Haisjackl, C.:
\newblock {Test Driven Modeling meets Declarative Process Modeling – A Case
  Study}.
\newblock Master's thesis, University of Innsbruck (2012)

\bibitem{Zug+12c}
Zugal, S., Haisjackl, C., Pinggera, J., Weber, B.:
\newblock {Empirical Evaluation of Test Driven Modeling}.
\newblock IJISMD (to appear, available online)

\bibitem{Kha+06}
Khatri, V., Vessey, I., Ramesh, P.C.V., Park, S.J.:
\newblock {Understanding Conceptual Schemas: Exploring the Role of Application
  and IS Domain Knowledge}.
\newblock Information Systems Research \textbf{17} (2006)  81--99

\bibitem{Fah+09a}
Fahland, D., Mendling, J., Reijers, H.A., Weber, B., Weidlich, M., Zugal, S.:
\newblock {Declarative versus Imperative Process Modeling Languages: The Issue
  of Understandability}.
\newblock In: Proc. EMMSAD '09. (2009)  353--366

\bibitem{Kahneman_2002}
Kahneman, D.:
\newblock Maps of bounded rationality: A perspective on intuitive judgment and
  choice.
\newblock Nobel prize lecture \textbf{8} (2002)  449--489

\bibitem{Zug+12b}
Zugal, S., Pinggera, J., Reijers, H., Reichert, M., Weber, B.:
\newblock {Making the Case for Measuring Mental Effort}.
\newblock In: Proc. EESSMod '12. (2012)  37--42

\bibitem{Zug+11}
Zugal, S., Pinggera, J., Mendling, J., Reijers, H., Weber, B.:
\newblock {Assessing the Impact of Hierarchy on Model Understandability-A
  Cognitive Perspective}.
\newblock In: Proc. EESSMod '11. (2011)  123--133

\bibitem{costain2007cognitive}
Costain, G.F.:
\newblock Cognitive Support During Object-oriented Software Development: The
  Case of UML Diagrams.
\newblock PhD thesis, University of Auckland (2007)

\bibitem{Bec+00}
Becker, J., Rosemann, M., Uthmann, C.:
\newblock {Guidelines of Business Process Modeling}.
\newblock In: Business Process Management, Models, Techniques, and Empirical
  Studies, London, UK, Springer-Verlag (2000)  30--49

\bibitem{MVDA+08}
Mendling, J., Verbeek, H., van Dongen, B., van~der Aalst, W., Neumann, G.:
\newblock Detection and prediction of errors in epcs of the sap reference
  model.
\newblock DKE \textbf{64} (2008)  312--329

\bibitem{MeRA10}
Mendling, J., Reijers, H.A., van~der Aalst, W.M.P.:
\newblock Seven process modeling guidelines (7pmg).
\newblock Information {\&} Software Technology \textbf{52} (2010)  127--136

\bibitem{BJM06}
Burton-Jones, A., N.Meso, P.:
\newblock Conceptualizing systems for understanding: An empirical test of
  decomposition principles in object-oriented analysis.
\newblock Information Systems Research \textbf{17} (2006)  38--60

\bibitem{Rej+11}
Reijers, H., Mendling, J., Dijkman, R.:
\newblock Human and automatic modularizations of process models to enhance
  their comprehension.
\newblock Inf. Systems \textbf{36} (2011)  881--897

\bibitem{Mood04}
Moody, D.L.:
\newblock {Cognitive Load Effects on End User Understanding of Conceptual
  Models: An Experimental Analysis}.
\newblock In: Proc. ADBIS '04. (2004)  129--143

\bibitem{Cru+07}
Cruz-Lemus, J.A., Genero, M., Morasca, S., Piattini, M.:
\newblock {Using Practitioners for Assessing the Understandability of UML
  Statechart Diagrams with Composite States}.
\newblock In: Proc. ER Workshops '07. (2007)  213--222

\bibitem{KHH00}
Kim, J., Hahn, J., Hahn, H.:
\newblock How do we understand a system with (so) many diagrams? cognitive
  integration processes in diagrammatic reasoning.
\newblock Information Systems Research \textbf{11} (2000)  284--303

\bibitem{Web+09b}
Weber, B., Reijers, H.A., Zugal, S., Wild, W.:
\newblock {The Declarative Approach to Business Process Execution: An Empirical
  Test}.
\newblock In: Proc. CAiSE '09. (2009)  270--285

\bibitem{ZuPiWe2012Toward}
Zugal, S., Pinggera, J., Weber, B.:
\newblock {Toward Enhanced Life-Cycle Support for Declarative Processes}.
\newblock Journal of Software: Evolution and Process \textbf{24} (2012)
  285--302

\bibitem{Pin+12a}
Pinggera, J., Furtner, M., Martini, M., Sachse, P., Reiter, K., Zugal, S.,
  Weber, B.:
\newblock {Investigating the Process of Process Modeling with Eye Movement
  Analysis}.
\newblock In: Proc. ER-BPM '12. (2013)  438--450

\end{thebibliography}

\end{document}